\documentstyle[12pt]{article}
\author{Hans - J\"urgen Schmidt}
\title{Stability and Hamiltonian formulation of higher 
derivative theories}
\date{}
\begin{document}
\maketitle

\bigskip

\centerline{
Universit\"at  Potsdam, Mathematisch-naturwiss. Fakult\"at}
\centerline{  Projektgruppe 
Kosmologie}
\centerline{
      D-14415 POTSDAM, PF 601553, Am Neuen Palais 10, Germany}

\bigskip

 Phys. Rev. D 49 (1994)  6354; 
Erratum Phys. Rev. D 54 (1996) 7906.

\bigskip

\begin{abstract}
We  analyze  the  presumptions which  lead  to  instabilities  in
theories  of order higher than second.  That type of fourth order
gravity  which leads to an inflationary (quasi de Sitter)  period
of  cosmic  evolution  by inclusion of one curvature squared term
 (i.e.  the Starobinsky model) is used as an example.  
The corresponding Hamiltonian formulation (which is necessary for
deducing  the 
Wheeler  de  Witt  equation) is found both  in  the  Ostrogradski
approach and in another form. 
  As an example,  a closed form solution of  the 
Wheeler  de  Witt  equation 
for a spatially flat Friedmann model and $L=R\sp 2$ is found.
The  method  proposed by Simon to bring fourth order  gravity  to
second order  can 
be (if suitably generalized) applied to bring sixth order gravity
to second order.
\end{abstract}

\bigskip

general mathematical aspects; 
 9880 Cosmology; 0450 Other theories of gravitation

\bigskip

\section{ Introduction }

It is quite general belief that curvature squared terms, if added

to  the  Einstein  - Hilbert  action,   describe  semiclassically
quantum corrections to General Relativity.  Further,  there is no
doubt  that the existence of an inflationary period  (exponential
expansion  of the cosmic scale factor) solves a lot  of  problems
connected  with the standard big bang model of the  universe.  So
there  is  no  wonder,  that the  Starobinsky  model  - curvature
squared  terms  lead  automatically to the  desired  inflationary
period - enjoyed so much interest in recent years. Now, Simon and
others 
formulated  some reasons speaking against the Starobinsky  model,
the  main  reason is the fact that the field equation  underlying
the Starobinsky model is of fourth order.
      
It  is the aim of the present paper to analyze  those  arguments,
which   are  connected  with  higher  (  =  higher  than  second)
derivative theories. 
The 80-pages article [1] entitled ``The problem of nonlocality in
string theory'' discusses in its sct. 2 the ``fundamental 
problems 
of   nonlocality    through   the   higher   derivative  
limiting 
procedure''.  The principal result of its subsection 2.1. is that
at least $ N - 1 $ of the solutions of a nondegenerate theory of 
order  $ 2N $ carry negative energy.  Eliezer and Woodard write:
  ``The energy is 
therefore unbounded below for all nondegenerate,  higher
derivative 
theories''.  This leads to the instability observed in almost all
fourth  and higher order theories.  The Starobinsky  cosmological
model  follows  from fourth order gravity,  and so it seems to be

a 
candidate for such an unstable theory, see e.g. ref. [2] which is
entitled   ``No   Starobinsky  inflation   from   self-consistent
semiclassical gravity''. 
We analyze that part of the arguments which  is 
connected with the higher order.  To  this  end  we 
specialize  the  Ostrogradski approach [3] (which is a method  to
bring  a  higher  order Lagrangian into Hamiltonian  form  - more
recent work on this topic can be found in ref. [4]) 
 to fourth order theories  in 
sct. 2 and give some intuitive examples. In sct. 3 we discuss the
question, whether fourth order theories lead to a minimum or only
to a saddle point of the action.  In sct.  4,  a method different
from  Ostrogradski's  one  is  proposed  to  bring  fourth  order
equations in a Hamiltonian form.
 
Then  we  are prepared to consider the Starobinsky model  [5]  in
sct.  5.  The  main problem comes from the $R\sp 2$-term,  so  we
simplify in sct.  5.1. by discussing the high-curvature limit and
derive  the corresponding Wheeler de Witt equation by the  method
described in sct.  4.  
Sct.  5.2.  discusses the question of  the 
superfluous  degrees of freedom of fourth order gravity and  sct.
5.3. gives the Starobinsky model in form of a power series not
yet 
found in the literature.
Sct.  6  is on sixth and higher order gravity.  It is included to
show which kind of problems additionally appear,  if $L=R+c_0R\sp
2$-gravity  is  intended to be the $k=0$-truncation  of  a  power
series
\begin{equation}
L = R+\sum_{i=0}\sp kc_iR\Box\sp iR 
\end{equation}
We  look  for the Newtonian limit of that theory  and  generalize
Simon's approach [2] to this Lagrangian (1.1) truncated at $k=1$.

Sct. 7 discusses the results.
\bigskip
\bigskip
\section{Ostrogradski's method for a fourth order system}
\setcounter{equation}{0}
We  follow Ostrogradski [3] but use the notation
 published  in  ref.  [1] which is more familiar to  the  present
reader, 
and  we specialize always to fourth order theories  which  follow
from a nondegenerate Lagrangian of second order.  So we consider 
a  1-dimensional  point particle with position $ q(t) $  at  time
$t$.  A dot denotes $ \frac{d}{dt} $ and the Lagrangian is of the
type
\begin{equation}
L = L(q, \dot q, \ddot q)
\end{equation}
where  $ q \in I$,  $I \ne \emptyset $ being a connected open
subset 
of the space $  R$ of all reals, and $ \dot q, \ddot q $  are 
allowed to cover all the reals. The momentum $P _2$ is defined by
\begin{equation}
P _2 = \frac{\partial L}{\partial \ddot q}
\end{equation}
In [1], the Lagrangian $L$ is defined to be nondegenerate if this
eq. (2.2) can be solved for $ \ddot q$ which takes place, 
loosely 
speaking, iff $\frac{\partial P _2}{\partial \ddot q} \ne 0 $.
To  avoid discussions for the case that (2.2) can be solved,  but
not uniquely, we additionally require that
$$\frac{\partial P _2}{\partial \ddot q} = F(q, \dot q) $$ Under 
this circumstance the Lagrangian is nondenegerate if and only  if
$F$ does not have any zeroes, i.e., it is a map 
$ F:I \times   R \longrightarrow   R \verb|\| \{0 \} $.
This we shall presume in the following. Then the Lagrangian (2.1)
can be written as 
\begin{equation}
L = \frac{1}{2}(\ddot q)\sp 2 F(q, \dot q)
+ \ddot q \, G(q, \dot q) + K(q, \dot q)
\end{equation}
To avoid discussions of differentiability, we simply require the 
three functions $F \ne 0, G, K$ to be real analytic ones.  
Then eq. (2.2) becomes
\begin{equation}
P _2 = \ddot q \, F(q, \dot q)+ G(q, \dot q)
\end{equation}
and it can be uniquely inverted to
\begin{equation}
 \ddot q = \frac{P _2 - G(q, \dot q)}{ F(q, \dot q)}
\end{equation}
The Euler - Lagrange equation following from eq. (2.1) reads
\begin{equation}
\frac{\delta L}{\delta q} \equiv
 \frac{\partial  L}{\partial q} - \frac{d}{dt} \frac{\partial 
L}{\partial \dot q} + \frac{d \sp 2}{dt\sp 2} 
 \frac{\partial L}{\partial \ddot q}=0
\end{equation}
Subsequently we write $q \sp{(0)} = q $ and 
$q  \sp{(n+1)}  =  \dot q \sp{(n)}  $.  Inserting  eq. (2.3) 
into 
eq. (2.6) we get an equation of the structure
\begin{equation}
0 = q \sp{(4)} F(q \sp{(0)},q \sp{(1)}) 
+ J(q \sp{(0)},q \sp{(1)},q \sp{(2)},q \sp{(3)}) 
\end{equation}
where $J$ is a real analytic function composed of $F, G$, and
$K$. 
From eq. (2.7) the notion of nondegeneracy becomes apparent: The 
second  order Lagrangian (2.1) is nondegenerate iff the Euler  - 
Lagrange  equation is a regular fourth order equation.  The  fact
that we restricted the domain of $q$ to the subset $I$ of $   
R   $  is  no  real  restriction  because  by  a  real   analytic
redefinition $ \tilde q(q) $ we could get $ \tilde I =   R$. 
To  get a Hamiltonian for this system,  one needs a  first  order
formulation with two coordinates. We define them as  
\begin{equation}
Q \sp 1 = q, \, Q \sp 2 = \dot q
\end{equation}
The two conjugate momenta are 
\begin{equation}
P _1 =  \frac{\partial L}{\partial \dot q} -
 \frac{d }{dt} \frac{\partial L}{\partial \ddot q}
\end{equation}
and  $P  _2  $ defined by  eqs.(2.2)/(2.4).  The  Hamiltonian  
$ H = H(Q \sp 1, Q \sp 2 , P _1, P _ 2) $ is 
obtained via
\begin{equation}
H = - L +  \sum \sp 2 _{n=1} P _n \dot Q \sp n 
\end{equation}
With these definitions the canonical equations
\begin{equation}
\dot Q \sp n =  \frac{\partial H}{\partial P _n}, \qquad 
\dot P _n = -  \frac{\partial H}{\partial Q \sp n}
\end{equation}
 take place, and their validity implies the validity of the Euler
- Lagrange equation (2.6). Now the arguments of $F, G, K$ are $ (
Q \sp 1, Q \sp 2 ) $ and we get
\begin{equation}
\dot Q \sp 1 =  Q \sp 2, \, \dot Q \sp 2 = (P _2 - G)/F
\end{equation}
and after some calculus
\begin{equation}
H = P _1 Q \sp 2 + \frac{1}{2F} (P _2 - G ) \sp 2 - K
\end{equation}
where  $ dH/dt = 0 $ follows from eq. (2.6).  So, $ H$ can be
called 
the energy of the system.  The essential point is that the energy
is  unbounded both below and above.  This is directly  seen  from
eq. (2.13) because $H$ is a linear function in $P _1 $.  
\bigskip
{\small {\it Remark}: In ref. [1], it was argued that the problem
lies 
in  the  fact  that energy is unbounded below.  More  exactly 
one 
should  say:  unbounded  below and  above;  supposed,  energy  is
unbounded  below  and bounded above,  then we simply  change  the
signs of both $L$ and $H$ and get the energy bounded below.  This
is  possible  because no sign of $H$ is preferred a  priori  - in
contrast to classical mechanics where the sign of $H$ is  defined
by the condition that kinetic energy is non-negative. }
\bigskip
\bigskip
$H =const.$ represents a first integral of eq. (2.6). It is, as
it 
must be the case,  a third order equation for $q(t)$,  and it has
the structure 
\begin{equation}
H = - q\sp{(1)}  q\sp{(3)} F + lower \,  order \, terms
\end{equation}
  There  is  a  singular  point at $  \dot  q  =  0$.  The 
definition eq. (2.13) of $H$ is essentially (i.e. up to
invertible 
linear  transformations  of $L$ and $H$ which do not  change  the
dynamics)    unique,     because    time-independent    canonical
transformations do not change it.  That $H$ is unbounded both
below 
and above can be seen from eq.  (2.14): Fixing the initial values
$q$,  $\dot  q  \ne 0$,  $\ddot q$,  one can freely  choose  $q 
\sp{(3)}$ (cf. eq. (2.7)) and gets $H$ as unbounded.
The  instability  following  from $H$ being unbounded  below  and
above  can  be described as follows.  {\bf A}:  In  the  particle
picture  one  gets  particles with positive  and  particles  with
negative energy.  Then the unlimited production of pairs of  such
particles is not prevented by energy conservation.   {\bf B}:  In
the  four-parameter  set  of solutions of  the  Euler  - Lagrange
equation  one  gets a subset of dimension $ \ge 1$  of  solutions
with negative energy. Let us make this last point more explicit. 
To this end we first consider what happens if we add such a total
derivative  to the Lagrangian that the functional dependence does
not change. This is done by 
\begin{equation}
\tilde L = L + \frac{d}{dt} [M - E \, t ]
\end{equation}
where $E$ is a constant and $M$ depends on $q$ and $ \dot q$. 
One gets 
\begin{equation}
\tilde P _1 = P _1 + \frac{\partial M}{\partial q}, \,
\tilde P _2 = P _2 + \frac{\partial M}{\partial \dot q}
\end{equation}
so that the condition of invertibility of $P_2$ does not  change,
and
\begin{equation}
\tilde F = F, \, \tilde H = H + E
\end{equation}
So  this  transformation,  too,  does not change  the  properties
discussed.
Let us continue with the discussion of negative energy solutions.
We  assume  that $q=0$ is a solution,  and we fix $E$  such  that
$q=0$ is a zero-energy solution. 
\bigskip
{\small {\it Remark}: In the first order Lagrangian $ L =
\frac{1}{2} \dot  q 
\sp 2 -  \frac{A}{2}  q \sp 2$ one has $p = \dot q$ and
$H  =  \frac{1}{2} p \sp 2 +  \frac{A}{2}  q \sp 2$.  For  $A=1$,
one  has  the  solutions $q=\sin t$ and $q=\cos t$  which  both 
have 
energy $H = \frac{1}{2}$.   For  $A=-1$, however, 
one  has  the  solutions $q=\sinh t$ and $q=\cosh t$  which  have
energy  $H = \frac{1}{2}$ and  $H =- \frac{1}{2}$ resp.  They sum
up to the zero energy solution $q=\exp t$.  This  is 
the instability meant. }
\bigskip
\bigskip
For the second order Lagrangian we consider only the terms up to 
second degree in the arguments. The terms $\dot q$,
$\ddot q$, $\dot q \ddot q$, $ q \dot q$ and
 $  q \ddot q +  \dot q \sp 2$ represent total  derivatives,  and
we use them to bring the general form to 
\begin{equation}
L = \frac{1}{2} \ddot q \sp 2 +
 \frac{A}{2} \dot q \sp 2 -  \frac{B}{2}  q \sp 2
\end{equation}
The corresponding Hamiltonian becomes
\begin{equation}
H = \frac{1}{2} \ddot q \sp 2 +
 \frac{A}{2} \dot q \sp 2 +  \frac{B}{2}  q \sp 2
 - \dot q \, q \sp{(3)}
\end{equation}
The Euler - Lagrange equation reads
\begin{equation}
0 =  q \sp{(4)} -
A \ddot q  - B  q
\end{equation}
The momenta are $$ P_1 = A \dot q - q \sp{(3)},  \, P_2 = \ddot q
$$.
For $A = B = 0$ one gets the positive energy solution $q=t \sp 2$
and the negative energy solution  $q=t \sp 3 + t$ . 
For $B = 0, \, A = \pm 1$, the solution 
$$q = \alpha t + \beta s(t) + \gamma c(t) $$ where 
$s(t)  =  \sinh  t$ for $A = 1$,  $s(t) = \sin t$ for $A  =  -1$,
analogously $c(t)$, has the energy $$
H = \pm \frac{1}{2}( \alpha \sp 2 - \beta \sp 2) + \frac{1}{2}
\gamma \sp 2 $$.
For $A = 0, \, B = 1$, the general solution of eq. (2.19) reads
$$q = \alpha  \sin t + \beta \cos t + \gamma \sinh t
+ \delta \cosh t $$
One gets 
$$H = \alpha  \sp 2  + \beta  \sp 2  - \gamma  \sp 2 
+ \delta  \sp 2  $$
The  general  case shows similarly that both signs  of  the 
energy 
appear. For non-linear equations, of course, the solutions do not
simply  add,  but  the  behaviour of the signs of the  energy  is
similar. 
\bigskip
\bigskip
\section{Minimal, not only stationary action}
\setcounter{equation}{0}
Instability  may occur,  if the action is not minimal,  but  only
stationary. We shall check, whether this type of instability can 
occur  in  the second order Lagrangian discussed.  More  on  this
topic,  especially applied to fourth order gravity,  can be found
in  ref.  [6];  however,  the point here is only to convince  the
reader   that  requiring  minimality  of  the  action  does   not
trivially rule out fourth order theories.
To do so, we develop the action $S[q+ \epsilon h]$ defined by
\begin{equation}
S[q] = \int _0 \sp T L \, dt
\end{equation}
into powers of $\epsilon$,  
where $L$ is the same as in eq. (2.1), and $T>0$. Without loss of
generality, the initial point of time was put $t=0$. 
Let $h$ be any differentiable function fulfilling
$h(0)=\dot h(0)=h(T)=\dot h(T)=0$.  After partial integration and
use of the notation eq. (2.6) we get
\begin{equation}
S[q+\epsilon h] =S[q]+ \epsilon \int _0 \sp T h 
\frac{\delta L}{\delta q} \, dt + \frac{\epsilon \sp 2 }{2}
V[q,h] + O(\epsilon \sp 3 )
\end{equation}
where 
$$
V[q,h] =  \int _0 \sp T
h \sp 2 \frac{\partial \sp 2 L}{\partial q \sp 2} +
 \dot h \sp 2 \frac{\partial \sp 2 L}{\partial \dot  q \sp 2} +
 \ddot h \sp 2 \frac{\partial \sp 2 L}{\partial \ddot q \sp 2} +
2h \dot  h \frac{\partial \sp 2 L}{\partial q \partial \dot q} +
2h \ddot h \frac{\partial \sp 2 L}{\partial q \partial \ddot q} +
2 \dot h \ddot h \frac{\partial \sp 2 L}{\partial \dot q \partial
\ddot q} 
dt $$
In dealing with $V$,  partial integration does not help.  So  one
should it discuss directly.
\bigskip
{\small  {\it  Remark:} Before discussing the fourth order  case,
let us repeat the behaviour for the harmonic oscillator
$L = \frac{1}{2} \dot q \sp 2 - \frac{1}{2} q \sp 2$. One gets
$$ V =   \int _0 \sp T \dot h \sp 2 -  h \sp 2 \, dt $$
One  only needs the boundary conditions $h(0)=h(T)=0$.  From  the
first  glance,  $V$ seems to possess only a saddle point,  but  a
Fourier analysis with 
$$ h = \sum _{n=1} \sp{\infty} a_n \sin(n \pi t/T) $$
gives
$$ V = \frac{T}{2} \sum _{n=1} \sp{\infty} 
(\frac{\pi \sp 2 n \sp 2}{T \sp 2}-1) a_n \sp 2 $$
For  $T<\pi$,  this  is positive  definite,  for  $T=\pi$,  it 
is 
positive  semidefinite,  and  only  for $T > \pi$ it  becomes  a 
saddle  point.  The  harmonic oscillator is {\it the  }  standard
example of a stable model, so one should require the action to be
locally minimal,  i.e.,  minimal if considered over  sufficiently
short but finite time intervals.}
\bigskip
\bigskip
Let us now come to the Lagrangian eq.  (2.18).  Inserting it into
eqs. (3.1), (3.2) we get
\begin{equation}
 V =   \int _0 \sp T \ddot h \sp 2+A \dot h \sp 2 -B  h \sp 2
\,dt
\end{equation}
We perform the same Fourier analysis than in the previous example
and get: 
The maximally allowed value $T$ depends on $A$ and $B$, but it is
always  positive,  so  that one has the same kind  of  stability 
here: If the time interval considered is sufficiently short, then
each stationary point of the action represents a minimum. 
\bigskip
\bigskip
\section{Another Hamiltonian formalism}
\setcounter{equation}{0}
Besides Ostrogradski's approach discussed in sct. 2, there exists
another  possibility  to  get a Hamiltonian from a  higher  order
Lagrangian. It has the advantage that the relation from classical
mechanics
\begin{equation}
p _i = \frac{\partial L}{\partial \dot q \sp i}
\end{equation}
remains  valid,  whereas Ostrogradski changed it  to  eq.  (2.9).

Further  possibilities to get a Hamiltonian are discussed in ref.
[7] (see also the references cited there), the difference is that
in  [7],  there  is always a constraint,  whereas we look  for  a
method where no additional constraint must be introduced. 
  
Again,  we start from eq. (2.1) and concentrate on Lagrangians of
the  type (2.3).  The difference is now that the new  coordinates
are chosen to be
\begin{equation}
q \sp 1 = q, \qquad q\sp 2 = \ddot q
\end{equation}
So the dependence of $ L(q\sp 1,  \dot q\sp 1,  q\sp 2,  \dot
q\sp 
2)$ is unique, whereas by use of  
eq.  (2.8), there is an ambiguity between $\dot Q \sp 1$ and
$Q\sp 
2$.
 To  show  up  the  procedure  we find  it  more  appropriate  to
concentrate on one single Lagrangian;  the general procedure
might 
become clear from it.  Moreover, it is just that Lagrangian which
will appear in sct. 5.  So we take
\begin{equation}
L = ( \ddot q ) \sp 2 e \sp{3q}
\end{equation}
To  prevent  an identical vanishing of $p_2$  according  to  eqs.
(4.1),  (4.2),  (4.3),  we add a suitable total derivative to the
Lagrangian (4.3). Let us first take 
\begin{equation}
\tilde  L  = L + \frac{4}{3} \,  \frac{d}{dt}[( \dot q) \sp 3   e
\sp{3q} ]
\end{equation}
i.e., an equation to be used in sct. 5.1.
\begin{equation}
\tilde  L = [(\ddot q) \sp 2 + 4 \ddot q( \dot q) \sp2 + 4(  \dot
q) \sp 
4 ] e \sp{3q}
\end{equation}
In a second step we take
\begin{equation}
\hat L =   L - 2 \, \frac{d}{dt}( \dot q  \ddot q  e \sp{3q} )
\end{equation}
i.e.,
\begin{equation}
\hat L = - [ (\ddot q) \sp 2 + 2  \dot q  q\sp{(3)}  + 6( \dot q)
\sp 
2 \ddot q ] e \sp{3q}
\end{equation}
We insert eq. (4.2) into eq. (4.7) and get
\begin{equation}
\hat L = - [(q \sp 2)\sp 2 + 2 \dot q\sp{1} \dot q \sp 2  + 6
(\dot q \sp 
1 ) \sp 2  q \sp 2 ] \exp(3q \sp 1)
\end{equation}
where we use $ \dot q \sp n \equiv \frac{d}{dt} (q \sp n ) $.
Applying (4.1) we get
\begin{equation}
p_1  = - [ 2 \dot q \sp 2 + 12 \dot q \sp 1 q \sp 2] \exp(3q  \sp
1), \qquad  p_2  = - 2 \dot q \sp 1 \exp(3q \sp 1)
\end{equation}
The Jacobian is
\begin{equation}
\frac{\partial(p _1,p _2)}{\partial(\dot q \sp 1,\dot q \sp 2) }
= - 4  \exp(3q \sp 1)
\end{equation}
This differs from zero, and so we can invert eq. (4.9) to 
\begin{equation}
\dot q \sp 1 = - \frac{1}{2} p_ 2 \exp(-3q \sp 1), \qquad
\dot q \sp 2 = [3 p _2 q \sp 2 - \frac{1}{2} p_1 ] \exp(-3q \sp
1)
\end{equation}
We  get from eqs.  (4.8)/(2.10) with the help of eqs.  (4.11) the
following Hamiltonian
\begin{equation}
H = - \frac{1}{2}(p_1p_2 - 3p_2 \sp 2 q \sp 2  ) \exp(-3q \sp 1)
+ ( q \sp 2 ) \sp 2  \exp(3q \sp 1)
\end{equation}
It  is essential to observe that the equation $\ddot q \sp 1 =  q
\sp  2  $  follows from the canonical  equations  of  $H$  (4.12)
without imposing it as  additional constraint.
So the equations (4.2) become automatically compatible.
 
\bigskip
To  give  some  feeling for Hamiltonians  with  negative  kinetic
energy  we  give  six typical examples - hoping that  this  gives
better insight than general formulations do. 
{\it Example  1}  Let $k$ be a parameter  fulfilling  $0<k<1$. 
We 
consider the Lagrangian 
\begin{equation}        
L = ( \ddot q) \sp 2 - 2( \dot q) \sp 2 + k q \sp 2
\end{equation}
for a 1-dimensional point particle $q(t)$.  The Euler  - Lagrange
equation reads
\begin{equation}        
0 = q \sp{(4)} + 2 \ddot q + kq
\end{equation}
We insert the ansatz
\begin{equation}        
q = e \sp{\lambda t} , \quad  0 = \lambda \sp 4
+ 2 \lambda \sp 2 + k 
\end{equation}
into eq. (4.14) which leads to
\begin{equation}        
\lambda = \pm \ i \ \sqrt{1 \pm \sqrt{1-k}}
\end{equation}
representing four different purely imaginary numbers.  Therefore,
the general solution of eq. (4.14) can be written as
\begin{equation}        
q(t) = \sum _{n=1} \sp 2  \quad c_n \sin(
t_n + t \sqrt{1 + (-1)\sp n \sqrt{1-k}})
\end{equation}
where $c_1, \, c_2, \, t_1, \, t_2 \, $  are the four integration
constants. Each solution is boun-ded in time. In the limiting
case 
$k=0$,  the  unbounded  function $q(t)=t$ is a solution.  In  the
other limiting case 
$k=1, \quad q(t) = t \sin t $ is also an unbounded solution.  
\bigskip
{\it Example  2}  Let $\epsilon $ be a parameter fulfilling  $0 
< 
\epsilon < 1$. Let
\begin{equation}        
H=  \frac{1}{2} p \sp 2 +  \frac{1}{2}
q \sp 2 + \frac{1}{2} P
 \sp 2 + \frac{1}{2}
Q \sp 2 + \epsilon q Q 
\end{equation}
be a Hamiltonian for two 1-dimensional point particles $q,  \, Q$
(or,  equivalently,  one  2-dimensional particle with coordinates
$(q,  \,  Q)$ );  $p$ is the momentum corresponding to $q, \, P$ 
to $Q$.  Because of the restriction put on $\epsilon ,  \,  H$ is
positive definite in all its arguments.  For $\epsilon = 0$, this
is nothing but two independent harmonic oscillators of  frequency
1. Only the term $ \sim \epsilon $ introduces some interaction. 
The canonical equations following from eq. (4.18) are
\begin{equation}        
\frac{\partial H}{\partial p} = \dot q = p
, \quad
\frac{\partial H}{\partial q} = - \dot p = q + \epsilon \,  Q = -
\ddot q
\end{equation}
\begin{equation}        
\frac{\partial H}{\partial P} = \dot Q = P 
, \quad
\frac{\partial H}{\partial Q} = - \dot P = Q + \epsilon \,  q = -
\ddot Q 
\end{equation}
To integrate the system it proves useful to eliminate $Q$ by
 use of eq. (4.19) as follows
\begin{equation}        
Q = - \frac{1}{\epsilon} \, (\ddot q + q)
\end{equation}
This leads with eq. (4.20)  to
\begin{equation}        
0 = q \sp{(4)} + 2 \ddot q + (1 - \epsilon \sp 2 ) q
\end{equation}
With $k = 1 - \epsilon \sp 2$ we meet exactly the system (4.14) 
from 
example 1.  The result of example 1 is in agreement with the KAM-
theorem  which  applies to the system considered here and  states
that there exists an interval of positive $\epsilon$-values, such
that  the  corresponding system is solved  by  torus-like  (i.e.,
periodic)  solutions.  Each  arbitrarily small  value  $\epsilon$

gives rise to a bifurcation of the frequency according to $ \vert
\lambda  \vert  $ in example 1.  Each solution is  periodic  and,
hence, bounded. In the limiting case $\epsilon \longrightarrow 0$
 corresponding  to  $k  \longrightarrow 1$,  the  equivalence  of
example 2 to example 1 breaks,  because for  $\epsilon = 0$,  all
solutions remain bounded here.  That this equivalence breaks as 
 $\epsilon = 0$, becomes also plausible from the relation (4.21)
between 
$Q$ and $q$. 
$H$  can  be considered to be the energy of the  system.  Let  us
express it as function of $q$ and its derivatives
 alone: (Calculations have been done 
by REDUCE 3.41)
\begin{equation}        
2H \epsilon \sp 2 = q [q + 2 \ddot q](1 -  \epsilon \sp 2)
+ [q \sp{(3)}] \sp 2 +
2 \dot q  q \sp{(3)} + (\ddot q) \sp 2  
 +( \dot q) \sp 2 (1 +  \epsilon \sp 2)
\end{equation}
A direct calculation leads to
\begin{equation}        
  \epsilon \sp 2 \frac{dH}{dt} = 
\dot Q [ q \sp{(4)} + 2 \ddot q + (1 - \epsilon \sp 2 ) q]
\end{equation}
hence, $H = const.$ follows from the $q$-equation (4.22), but 
$H = const.$ implies a solution for $\dot Q \ne 0 $ only.
Let  us  mention that the fourth order equation of example  1  is
equivalent to the positive definite Hamiltonian of example 2. 
\bigskip
{\it Example  3} Let us start again from the system of example 
1, 
but  now we apply the Ostrogradski approach to make a Hamiltonian
from it.  New coordinates are $Q \sp 1 = q, \, Q \sp 2 = \dot q$,
see eq. (2.8), new momenta are 
\begin{equation}        
 P _1 = \frac{\partial L}{\partial \dot q} - 
 \frac{d}{dt}
 \frac{\partial L}{\partial \ddot q}
= - 4 \dot q - 2 q \sp{(3)}
, \quad P _2 = \frac{\partial L}{\partial \ddot q } = 2 \ddot q
\end{equation}
see eqs. (2.9), (2.2). 
The Hamiltonian is
\begin{equation}        
\tilde H = - L + \sum _{n=1} \sp 2  \quad P_n \dot Q \sp n 
= P_1 Q \sp 2 + \frac{1}{4} P_2 \sp 2 + 2 (Q \sp 2) \sp 2
- k (Q\sp 1) \sp 2 
\end{equation}
Now, let us forget about the origin of       
$ \tilde H $ and calculate the canonical equations. Inserting 
 $Q \sp 1 = q$, we get after some calculus $Q \sp 2 = \dot q,
 \,  P_1  = - 4 \dot q - 2 q \sp{(3)},  \,  P_2 = 2 \ddot q $ and
finally 
\begin{equation}        
0 = q \sp{(4)} + 2 \ddot q + kq
\end{equation}
i.e., as it should be, just  eq. (4.14) of example 1. 
We insert the values for $P_n,  \,  Q \sp n$ onto $\tilde H$  and
get
\begin{equation}        
\tilde H = (\ddot q) \sp 2 - 2 \dot q  q \sp{(3)}
- 2 (\dot q) \sp 2 - k q \sp 2
\end{equation}
This is also a conserved quantity:
\begin{equation}        
\frac{d \tilde H}{dt} = - 2 \dot q \, 
[ q \sp{(4)} + 2 \ddot q + kq]
\end{equation}
but $ \tilde H $ is not evidently bounded. Clearly, each solution
remains bounded, but the set of solutions for a fixed value 
$ \tilde H $  need not to be bounded:  let us insert the solution
(4.17)  
of example 1 with $t_1 = t_2 = 0$, then one gets
\begin{equation}        
 \tilde H = 2 \epsilon \, [c_2 \sp 2
(1 + \epsilon) - c_1 \sp 2
 (1 - \epsilon)   ]
\end{equation}
 $  \tilde  H $ can take each real value,  and  with  arbitrarily
fixed  value   $  \tilde H $,  we can find an  unbounded  set  of
functions solving for just this  $  \tilde H $.
Let  us  compare this result with the analogous  calculations  in
example 2: There one gets from the same initial conditions
\begin{equation}        
 H = c_2 \sp 2
(1 + \epsilon) + c_1 \sp 2
 (1 - \epsilon) 
\end{equation}
Up  to  the inessential prefactor $2 \epsilon $,
  it is just the other sign  in 
front of $c_1 \sp 2$ which makes the difference.  Here, $H \ge
0$, and 
the set of solutions for a fixed value $H$ forms a compact set 
of 
bounded functions,  moreover:  it is a uniformly  bounded set of 
functions. So the conserved quantity   $ \tilde H $ should not be
considered  as the energy of the system,  because   $ H $  better
meets  the  point.  This is another argument  against  using  the
Ostrogradski approach.  Let us further mention, that the Poisson
bracket 
of  these  two conserved quantities $H$, $ \tilde H$
 identically vanishes,  so  it 
does not give rise to a further conserved quantity.  

\bigskip
{\it Example 4} We take now the same
 $ \tilde H $ as in example 3, but we interchange coordinates and
momenta ($ P_1  \longrightarrow q, \, Q \sp 1  \longrightarrow 
-p, \, P_2 \longrightarrow Q, \, Q \sp 2 
 \longrightarrow -P 
$).
\begin{equation}        
\hat H = 2 P \sp 2 - k p \sp 2 - q \, P + \frac{1}{4} Q \sp 2
\end{equation}
Here,  the kinetic energy is indefinite, but the solutions 
are,  of course,  also only the periodic ones of example 1. Here,
we  see  that two different Hamiltonians may  describe  the  same
system,  one  has  indefinite,  the  other has  definite  kinetic
energy. On the other hand, a regular Hamiltonian must have a non-
vanishing Jacobian
\begin{equation}        
J =  \frac{\partial ( \dot q, \, \dot Q )}{\partial (p, \, P)}
=  \frac{\partial  \sp 2 H}{\partial p \sp 2  } \,
 \frac{\partial \sp 2 H }{\partial  P \sp 2  }
- ( \frac{\partial  \sp 2 H )}{\partial p \partial  P} ) \sp 2
\end{equation}
Here,  $J  <  0$. 
(This $J$ is the inverse to the Jacobian used in eq. (4.10); this
does not matter since only the sign of $J$ is essential here.)
 A one-parameter family  of  regular  canonical 
transformations connected with the identity transformation cannot
change the sign of $J$.  Therefore, the Hamiltonians $H$ (example
2) and $ \hat H$ (example 4) cannot be continuously deformed into
each other by such a transformation, because in example 2, $H$ is
positive definite,  and $J > 0$. Nevertheless, they describe the 
same system. 
\bigskip
{\it Example 5} Let us take the Hamiltonian
\begin{equation}        
\bar H = Pp + \frac{\epsilon }{2}(q\sp 2 + Q \sp 2) + qQ, \quad 
0 < \epsilon <1
\end{equation}
Both  the  kinetic and the potential part are indefinite. 
The canonical equations give
\begin{equation}        
P = \dot q, \quad p = \dot Q, \quad
\dot P = - q - \epsilon Q, \quad
\dot p = - Q -  \epsilon q
\end{equation}
After some calculus we get
\begin{equation}        
0 = q \sp{(4)} + 2 \ddot q + (1 - \epsilon \sp 2 ) q
\end{equation}
which is again the previously discussed system. 
\bigskip
{\it Example 6} Now we start from a Lagrangian which differs 
from 
example 2 only in two changes of a sign. 
\begin{equation}        
H \sp * =  \frac{1}{2} p \sp 2 +  \frac{1}{2}
q \sp 2 - \frac{1}{2} P
 \sp 2 - \frac{1}{2}
Q \sp 2 + \epsilon q Q, \quad 0< \epsilon <1 
\end{equation}
So,  both kinetic and potential energy are indefinite. But as was
seen in example 5, this does not exclude the equivalence. 
Let  us first consider the limiting case $ \epsilon =  0$.  Here,
again, it is fully  equivalent to the second example: There is no
interaction between the two oscillators, and there is no a priori
sign  preferred  for  the  energy.  $H$ and $H  \sp  *  $ are 
two 
conserved quantities, whose Poisson bracket vanishes.
The situation drastically changes if we come back to
 $  \epsilon > 0$.  One of the presumptions of the KAM-theorem is
no more valid, and so we expect qualitatively different solutions
for  arbitrarily  small values  $ \epsilon  $.  In  the  particle
picture  one can imagine the following:  The spontaneous creation
of pairs of particles, one with positive, the other with negative
energy,  is energetically allowed,  and it {\it should} take
place 
with a typical doubling time $ \sim 1/  \epsilon  $. 
For  fixed energy,  arbitrarily large momenta  are  possible.  We
perform  the calculations analogous to the previous ones.  We can
prevent  any  calculations  if we look at  $H$ and $H  \sp  *  $:
Multiplying $P$ and $Q$ by $i$ and multiplying $\epsilon $ by 
$(-
i)$,  one  is  changed into the other.  So,  clearly,  the  other
formulas are valid if  $\epsilon \sp 2 $ is replaced by
 $( - \epsilon \sp 2) $. Then the dynamics follows from
\begin{equation}        
0 = q \sp{(4)} + 2 \ddot q + kq, \quad
k =  1 + \epsilon \sp 2 
\end{equation}
and it is example 1 with $k > 1$.  The corresponding fourth order
polynomial for $ \lambda $ is then solved by 
\begin{equation}        
 \lambda = \pm i \sqrt{1 \pm i  \epsilon}
= \pm \frac{ \epsilon }{2} \pm i(1 + \frac{ \epsilon \sp 2}{8} )
+ O( \epsilon \sp 3)
\end{equation}
The  four  solutions correspond to the four combinations  of  the
signs "$\pm $". Therefore, the general solution can be written as
\begin{equation}        
q (t) = \sum _{n=1} \sp 2  \quad c_n \exp [
t (-1)\sp n ( \frac{ \epsilon }{2} + O( \epsilon \sp 3))]
\,  \sin[t_n + t(1 + \frac{ \epsilon \sp 2}{8}  + O( \epsilon \sp
4) ) ]
\end{equation}
where $c_1, \, c_2, \, t_1, \, t_2 \, $  are the four integration
constants. (By the way, the $ O( \epsilon \sp 3) = [
\frac{9}{64}  +  O( \epsilon \sp 2)] \epsilon \sp 3  $  for  this
formula.)  $q(t) \equiv 0$ is the only bounded solution,  and for
$c_2 \ne 0$,  one gets an exponential increase as expected.  This
is,  of course, a resonance effect. If, on the other side, $H \sp
*$ is altered by a suitable positive factor in front of $p\sp 
2$, 
then  for  small $\epsilon$,  the general solution remains to  be
periodic, but the periods are mixed.  
\bigskip
\bigskip
\section{The Starobinsky model}
\setcounter{equation}{0}
In ref. [5], Starobinsky proposed to use
\begin{equation}
L = ( \frac{R}{2} - \frac{l \sp 2}{12} R \sp 2 ) \sqrt{-g}
\end{equation} 
as gravitational Lagrangian.  Here,  $R$ is the curvature scalar,
$g$  the determinant of the metric of space-time,  and $l$  is  a
length being somehow in the region $l = 10 \sp{-28} cm$. 
\bigskip
\bigskip
\subsection{The high-curvature limit}
Let us first consider the high-curvature limit
\begin{equation}
\tilde L =  \frac{1}{36} \, R \sp 2  \sqrt{-g}
\end{equation} 
For the metric of a spatially flat Friedmann model 
\begin{equation}
ds \sp 2 = dt  \sp 2 - e\sp{ 2q(t)}(dx \sp 2+dy \sp 2+dz \sp 2)
\end{equation} 
we get 
\begin{equation}
R = - 6 \ddot q - 12( \dot q) \sp 2, \qquad g = - e\sp{6q}
\end{equation} 
and the Lagrangian becomes
\begin{equation}
\tilde L = [ \ddot q +2( \dot q) \sp 2] \sp 2  e\sp{3q}
\end{equation} 
which coincides with eq. (4.5).
Now we could apply both the Ostrogradski approach sct.  2 as well
as  the  approach  of sct.  4.  We prefer to use the  latter  one
because of the validity of eq. (4.1), but for comparison we write
down both of them.
Let  us  first  apply the Ostrogradski  approach  to  eq.  (5.5).
Looking  at eqs.(4.5),  (4.4),  (4.3) one can see that (up to  a 
divergence) we have to consider 
\begin{equation}
 L =  ( \ddot q ) \sp 2  e\sp{3q}
\end{equation} 
Applying the formalism of sct. 2 we get
$$ Q\sp 1 = q, \qquad  Q \sp 2 = \dot q, \qquad 
P_1=-2\frac{d}{dt}
( \ddot q  e\sp{3q}), \qquad   P_2= 2\ddot q  e\sp{3q} $$ and
then
\begin{equation}
H = P_1Q\sp 2 +  \frac{1}{4}  (P_2) \sp 2  \exp(-3Q\sp 1)
\end{equation} 
i.e.,
\begin{equation}
H =  e\sp{3q}[
 (\ddot q) \sp 2 - 2\dot q (
q\sp{(3)}+3\dot q \ddot q)]
\end{equation} 
As is to be expected,  the canonical equations to the Hamiltonian
$H$ eq. (5.7) give again the original system, where $H$ eq. (5.8)
represents  a conserved quantity.  Moreover,  one knows that  the
gravitational  field  equation  (here especially  its 
 zero-zero-component) forces $H$ to vanish.  This can easiest  be
shown by making the ansatz $dt\sp 2 = N\sp 2(\tau)d\tau\sp 2$ and
putting  $N=1$  only  after the variation (and not before  as  we
did). 
Let us now come to the Wheeler de Witt equation (i.e.,  the  zero
energy  Schr\"odinger equation of a cosmological model) for  this
system. In units where $\hbar =1$ it is obtained via substituting
$P_n$  by $i\partial _n \equiv 
i \frac{\partial}{\partial Q\sp n}$.  Applying this to 
eq.  (5.7) one can see that the fact that $P_1$ is contained
 linearly and not 
quadratically  gives as consequence that one of the  coefficients

of 
the  Wheeler  de Witt equation fails to be real.  A third  reason
speaking against this approach is the fact, that $Q\sp 2 =\dot q$
(which is just the Hubble parameter of the cosmological model eq.
(5.3)) is not invariantly defined, it changes its sign by a
change 
of the time direction. 
Now we try the same with the formulas of sct.  4. In fact, we can
work directly with eqs. (4.3), (4.6 - 4.12). If we re-insert eqs.
(4.2),  (4.9) into eq. (4.12) we get exactly the eq. (5.8). So no
second conserved quantity appears here,  and the Wheeler de  Witt
equation is derived as follows. 
The  material  from eq. (5.9) till eq.  (5.13) is taken from 
ref. 
[8], and from eq. (5.14) till eq. (5.15) is taken  ref. [9]. 
The Lagrangian $\hat L$ eq. (4.8) has the structure
\begin{equation}
\hat L = \frac{1}{2}
 g_{ij} \dot q\sp i\dot q\sp j - V(q\sp i)
\end{equation}
where  $g_{ij}$ depends on the $q\sp i$ only and the Einstein sum
convention is applied. One gets
\begin{equation}
p_i = \frac{\partial L}{\partial \dot q\sp i} = g_{ij} \dot q \sp
j
\end{equation}
To apply the Hamiltonian formalism, it is necessary to invert eq.
(5.10)  such  that  the velocities are written  in  dependence 
of 
coordinates and momenta. This is possible if and only if $g_{ij}$
is an invertible matrix. This takes place for the case considered
here: A comparison of (4.8) with (5.9) gives 
$$  g_{11} =-6q\sp 2\exp(3q\sp 1) ,\qquad 
  g_{12} = -\exp(3q\sp 1) ,\qquad   g_{22} =0$$
Let $g\sp{ij}$ be the inverse matrix to $g_{ij}$.  Then eq.
(5.10) can 
be inverted to 
\begin{equation}
\dot q \sp i = g\sp{ij} p_j
\end{equation}
In the interesting case (4.8) this gives
$$  g\sp{22} =6q\sp 2\exp(-3q\sp 1) ,\qquad 
  g\sp{12} = -\exp(-3q\sp 1) ,\qquad   g\sp{11} =0$$
The Hamiltonian becomes
\begin{equation}
H=p_i\dot q\sp i - L= \frac{1}{2}
 g\sp{ij} p_ip_j+ V(q\sp i)
\end{equation}
and here, $V=(q\sp 2)\sp 2 \exp(3q\sp 1)$. If we quantize now by
 substituting 
$p_n$ by $i \frac{\partial}{\partial q\sp n}$ then the  procedure
is no more covariant, and the factor ordering problem appears. In
classical mechanics this problem is absent, because $g_{ij}$ is a
constant matrix. We cirumvent the problem by
 substituting 
$p_n$  by  $i \nabla _n$ where $\nabla  _n$ denotes  the 
covariant 
derivative into $q\sp n$-direction
 with respect to the metric  $g_{ij}$. Then the Wheeler 
de Witt equation reads
\begin{equation}
0 = (\Box - 2V)\psi(q\sp i)
\end{equation}
where $\psi$ is the world function and 
$$ \Box = \nabla _i \nabla  \sp i =
\frac{1}{\sqrt{-g}}\partial _i \sqrt{-g} g\sp{ij}\partial_j
$$ is the D'Alembertian. For our example we get 
\begin{equation}
0  = [3 \partial_2 q\sp 2 \partial_2 - \partial_1   \partial_2- 2
(q\sp 2)\sp 2 \exp(6q\sp 1)]\psi
\end{equation}
To simplify, let us apply the following transformation
$$ \sigma=\exp(-3q\sp 1), \qquad \tau=\frac{3}{2} q\sp 2
\exp(3q\sp 1)$$
This transformation explicitly brings $g_{ij}$ to the flat form 
$g_{\sigma\tau} = 1$, 
$g_{\sigma\sigma} = 0$, 
$g_{\tau\tau} = 0$. The Lagrangian becomes
$$L = \frac{1}{\sigma}\dot \sigma \dot \tau- \tau \sp 2 \sigma$$
The Hamiltonian is correspondingly
$$H = \sigma[\pi_{\tau}\pi_{\sigma}+ \tau\sp 2]$$
and the Wheeler de Witt equation reads
\begin{equation}
0  = [\partial_{\tau}\partial_{\sigma}- \tau\sp  2]\psi(\sigma, 
\tau) \qquad .
\end{equation}
This linear differential equation can be solved in closed form by
\begin{equation}
 \psi = \int_{-\infty} \sp{\infty} a(\lambda)
\exp(\lambda \sigma + \frac{\tau \sp 3}{3 \lambda}) d\lambda
\end{equation}
where  the amplitude function $a$,  $a(0)=0$,  can be arbitrarily
chosen both as continuous as well as a sum of $\delta$-functions.

\bigskip
\bigskip
\subsection{Superfluous degrees of freedom}
We look at higher ( = higher than second) order gravity  theories
under the point of view that the higher order yields more degrees
of freedom than is to be expected. 
 Further results on higher order  gravity  and 
inflationary phase of cosmic evolution can be found in refs. [10
- 
24]  (a  list  which  is not intended  to  be  representative 
but 
essentially  contains  the papers we refer to in  the  subsequent
text). Let us start with some historical comments taken from ref.
[14].
 In [25],  Weyl proposes a new theory which is intended to  unify
gravitation  with  electromagnetism.   He  both  generalizes  the
Riemannian  geometry underlying the General Relativity theory  of
Einstein  to  a non-integrable theory (introduction of  the  Weyl
vector)  as  well as the Einstein - Hilbert Lagrangian (which  is
linear in curvature) to a Lagrangian quadratic in  curvature.  At
that point Weyl already realized, [25, p.  477]: "Dies hat zur 
Folge,   da\ss \  unsere   Theorie   wohl   auf   die  
Maxwellschen 
elektromagnetischen,    nicht    aber   auf   die   Einsteinschen
Gravitationsgleichungen    f\"uhrt;    an   ihre   Stelle   
treten 
Differentialgleichungen 4.  Ordnung." (Translation:  This has the
consequence, though our theory leads to Maxwell's electromagnetic
equations,  it  fails to lead to Einstein's  gravitational  ones;
instead  of  them,  4th order differential equations appear.)  We
cite  this sentence to show that already in 1918 the  possibility
of fourth order gravitational field equations had been  discussed
as alternative to General Relativity.  One year later, Pauli [26]
calculated  the static spherically symmetric solutions of  Weyl's
theory;  he  got the result that the Schwarzschild solution is  a
solution  for  all the equations following from one of the  three
Lagrangians
\begin{equation}
R \sp 2 , \  R _{ik} R \sp {ik} , \  R _{ijkl} R \sp {ijkl}
\end{equation}
($R $ is the curvature scalar, $ R _{ik} $    the Ricci tensor, 
$ R _{ijkl} $   the Riemann tensor.)
Pauli  assumed  the  Weyl vector to be zero,  so that  he  had  a
Riemannian structure of Lorentz signature as underlying geometry.
He concluded that measurements of Mercury perihelion advance  and
light  deflection in the field of the Sun which are in  agreement
with  General  Relativity  are also in  agreement  with  all  the
variants  of Weyl's theory,  but the fourth order theory has  too
many  ambiguities (both in finding the correct Lagrangian as well
as choosing the correct solution);  more explicitly this is  done
in [27].  Today one can say more generally:  Each vacuum solution
of  Einstein's equation is also a vacuum solution of each of  the
variants  of fourth order gravity (where Pauli [26] believed this
to be the case for the first two expressions in eq. (5.17) only)
.
Pauli [26] wrote about the superfluous degrees of  freedom,  that
they  are  a consequence of the fact that he only considered  the
vacuum equations and that it should be possible to cancel them by
finding the correct interior solution at the source.  The  latter
is  only a mathematical problem;  we proceed on this line in sct.
6.1.  Further,  he  assumes that the far-field of a mass $m$  can

be 
developed  into powers of $m/r$,  where $r$ is the distance from 
the 
center of the source. 
     The last point we want to repeat from the old papers is  the
 following:  $  R  $ has dimension $ <length> \sp {-2} $   ,  and
therefore, the action
\begin{equation}
\int R \sp m \sqrt{-g} \, d \sp n x
\end{equation}
(where  $  g  $  is the determinant of the metric in the 
$n$-dimensional space-time) is scale-invariant (i.e., does not
change 
by  a change of the used length-unit) if and only if $ n =  2m  $
holds. For the usual case $  n = 4$, this gives $  m = 2$, an 
argument which was already used by Weyl in 1918. 
  Now  let us come back to Simon's argument [2] that the 
superfluous 
degrees of freedom have to be cancelled: Surely, he has found one
possibility,  but  that  one  is a priori  not  better  than  the
following ones. In units where $ 8 \pi G = c = 1 $ we use the
Lagrangian 
\begin{equation}
L =( \frac{1}{2} R + \frac{1}{4} k \sp 2 C \sp 2 - \frac{1}{12}
 l \sp 2 R \sp 2 ) \sqrt{-g}
\end{equation}
where
\begin{equation}
C \sp 2 = C_{ijkl} C \sp{ijkl}
\end{equation}
is  the  square of the Weyl tensor and can be written  as  linear
combination of the terms in eq. (5.17).  The Lagrangian (5.19)
gives 
rise to a tachyonic-free theory if and only if both $ k \sp 2 \ge
0 $  and  $ l \sp 2 \ge 0 $  hold.  (In principle,  $ k \sp 2  $ 
and 
 $ l \sp 2  $ may have both signs, we prefer the nontachyonic
case.)
     It holds (see ref.  [28], and ref. [29] for the presentation
used here):
If we redefine the original metric $ g_{ij} $ to $ G_{ij} $   
via
\begin{equation}
 G_{ij} =  g_{ij} - 2 k \sp 2 R_{ij} +
\frac{1}{3} ( k \sp 2 - l \sp 2 ) R \, g_{ij}
\end{equation}
then  the linearized Einstein tensor of $ G_{ij} $   vanishes  if
and  only if     $ g_{ij} $   solves the linearized fourth  order

equation  following from   eq. (5.19).  For microscopically 
small 
lengthes  $ k $ and $ l $ both metrics cannot be distinguished by
experiment, and so one is free to use  $ G_{ij} $     as physical
metric possessing the required second-order dynamics, at least on
the linearized level.
 The second possibility is the following:  For the spatially flat
Friedmann  model the typical solution of fourth order gravity  is
(cf.   e.g.  refs. [19] and [20]):  Damped  oscillations  around 
the 
expansion law of the Einstein de Sitter model.  This is not  only
due to the high symmetry: In ref. [30] there is considered the
general 
anisotropic  Bianchi type I model,  and the result was the  same.
The general behaviour with inhomogeneous models is not known, but
there  exist  reasons (by the conformal transformation of  fourth
order  gravity  to  Einstein's theory with  a  minimally  coupled
scalar field,  see e.g.  ref. [31],  where flat space is related
to  a 
local minimum of the potential) to believe that there are similar
typical solutions.  We interpret them as follows: The superfluous
degrees  of freedom are just the phases of the oscillations,  and
by the damping of the amplitudes they simply disappear. 
\subsection{ Starobinsky inflation as power series}
In  this section we consider in more details than can be found in
the  literature:  In  which sense  the  Starobinsky  inflationary
solution can be developed in a power series ? To this end we make
the  following consideration (which makes more explicit what  has
been done in ref. [20],  sct. 5, especially [20], eq. (18)). It
is 
essential to note that there is neither a cosmological   term nor
an  additional  inflation or scalar field - all  inflation  comes
from the $ R\sp 2$ -term in the Lagrangian (5.19). The exact
inflationary 
de  Sitter  space-time  is defined by  eq.  (5.3)  with 
$q(t) = ht$, $h$ having  a 
constant positive value.  In general,  $h = \dot q$ is the Hubble
parameter. The quasi de Sitter stage  is  that 
period,  where   $  \vert  dh/dt  \vert  \ll  h\sp 2 $  . 
Now, the field equation 
following  from  Lagrangian (5.19) is considered.  The  Friedmann
model is conformally flat,  and so, the term with the Weyl tensor
identically vanishes. So, without loss of generality we put $k =
0 $. 
Further, we consider  the non-tachyonic case $ l\sp 2  > 0$ only.
  Suen [17, 18] showed 
 that flat space is unstable and that the Starobinsky solution is
stable.  A  partial  stability  of flat  space  with  respect  to
initially expanding perturbations can be found, however:
in [20] it was shown:  { \it All} vacuum solutions representing 
an expanding spatially flat Friedmann model of the field equation
following  from the Lagrangian (5.19) with $ l\sp 2  > 0 $ can 
be 
integrated  up  to infinity;  they all have the  same  asymptotic
behaviour:  Damped  oscillations with frequency $ 1/l$ about  the
Einstein de Sitter model $ q = t \sp{2/3} $ .  
The spatially flat  Friedmann 
model has the metric (5.3).
We take the Lagrangian (5.19) and get a fourth order 
differential 
equation  for  the metric.   However,  the 
 00-component of the differential equation is a constraint and 
gives 
a second order equation for $h$ which reads
\begin{equation}
2h\frac{d \sp 2 h}{dt \sp 2} - [ \frac{ dh}{dt}] \sp 2
+ 6 h \sp 2  \frac{ dh}{dt}  =  - h \sp 2 l\sp{- 2} 
\end{equation}
Linearization of this equation gives $ 0 = 0$, so that it is
clear 
that  the linearized equation gives no information about the full
one.  It holds:  Each solution of (5.22) is also a solution of
the 
other 9 components of the field equation;  however, each function
$h$  solves  the linearized equation  but in general  not  the 
linearized trace equation which reads simply
\begin{equation}
\frac{d \sp 3 h}{dt \sp 3} +  l\sp{- 2} \frac{ dh}{dt}  =  0 
\end{equation}
Next,  it is clear that eq. (5.22) has a singular point at $h =
0$, 
and so the numerical integration has to be done with  care.   The
best method to integrate the system numerically is the following:
One  uses the constraint only at the initial moment and then  one
integrates the trace equation;  the trace is regular even for $ h
= 
0  $.  Also  the existence of oscillations with frequency $  1/l$
becomes clear from (5.23), and the sign of the r.h.s. of eq.
(5.22) decides 
whether  the  oscillations are damped or not.  But the result  of
[20],  that for $ l\sp 2  > 0$ and initial value $h > 0$ the 
equation 
(5.22) can be integrated up to infinite time $t$,  is strong  and
does  not depend on the numerics,  and it does not change  if  we
include classical matter like dust or radiation. 
 In  eq. (5.22) the inflationary period 
 can be found by requiring that the first  two  items 
are  negligible in comparison with the third one.  (Afterwards it
will turn out that the first two terms remain finite whereas  the
third  one tends to infinity as $ h \longrightarrow \infty $;  so
this  approximation is consistent.) We get the first step of  the
approximation by removing the first two terms of  eq. (5.22); 
this 
leads to the equation
\begin{equation}
\frac{dh}{dt} = - \frac{1}{6 l \sp 2 }
\end{equation}
The  larger the value $ h$ ,  the better the approximation
(5.24). 
This justifies to use a Laurent sequence in $ h\sp 2 $ as general
ansatz as follows
\begin{equation}
\frac{dh}{dt} = - \frac{1}{l \sp 2 }
\sum \sp{ \infty } _{i=0} (-1) \sp i g_i ( h l ) \sp{-2i}
\end{equation}
We  included  such  powers of the length $l$ as  factors  that 
the 
coefficients $ g_i $ become real numbers.  Comparing (5.24)  with
(5.25) 
one gets $ g_0  = 1/6 $.  The motivation of the factor $ (-1) \sp
i $  will become clear afterwards:  all numbers $ g_i $ will 
turn 
out to be positive. Just for the same reason we did not write odd
powers  of  $  1/h $ in (5.25),  because  all  their 
coefficients 
automatically  vanish  if we insert the sequence  into  eq.
(5.22). 
This   is  very  satisfactory,   because  even  powers  of $ 
1/h$ 
correspond  to powers of $ \hbar $ whereas the odd  powers  would
correspond to $ \sqrt \hbar $ , which is a less natural quantity.
  The coefficients $g_i $  can be obtained as follows:  $ h $  
times    the derivative of eq. (5.25) gives
\begin{equation}
h \frac{d \sp 2 h}{dt \sp 2} = \frac{dh/dt}{l \sp 2}
\sum \sp{ \infty } _{i=0} (-1) \sp i  2i g_i (hl) \sp{-2i}
\end{equation}
Now we insert eqs. (5.25), (5.26) into (5.22),
 multiply by $ l\sp 2 $  and get 
step by step
\begin{equation}
 \frac{d h}{dt } [  6 h \sp 2 l \sp 2 +
\sum \sp{ \infty } _{i=0} (-1) \sp i (4i+1) g_i (hl) \sp{-2i}
] = - h \sp 2
\end{equation}
After division by $ (- h\sp 2 ) $ and some rearrangement we get
\begin{equation}
\sum \sp{ \infty } _{k=0} (-1) \sp k 6 g_k  (hl) \sp{-2k}
 - \sum \sp{ \infty } _{k=1} (-1) \sp k  (hl) \sp{-2k}
   \sum \sp{k-1 } _{i=0} (4i+1) g_i g_{k-i-1}  = 1
\end{equation}
The absolute value of eq. (5.28) gives again $  g_0   =  1/6 $ ,
and 
for each $ k > 0 $  we get
\begin{equation}
g_k = \frac{1}{6} \sum \sp{k-1 } _{i=0} (4i+1) g_i g_{k-i-1}
\end{equation}
e.g.  $  g_1 = 1/6 \sp 3  =  1/216, \,   g_2  = 1/6 \sp 4$, $ 
g_3  =  
65/6 \sp 7$.
The next natural step seems to be the insertion of (5.29) into
the 
ansatz  (5.25).  But  it turns out that one gets the  result 
more 
quick  by integrating that equation which is obtained from 
(5.27) 
after  division  by $ h \sp 2 $  (time-translation is only  a 
coordinate  transformation,  so we get no essential  constant  of
integration):
\begin{equation}
   6 l \sp 2 h  +
\sum \sp{\infty } _{i=0}
(-1) \sp{i+1}  g_i \frac{4i+1}{2i+1}
h \sp{-1} (hl) \sp{-2i}  = - t 
\end{equation}
This equation can be inverted as follows:
\begin{equation}
h = - \frac{t}{6l \sp 2}
- \frac{1}{6t} + \sum \sp{ \infty } _{i=1}
f_i \, (l/t) \sp{2i}  \, t \sp {-1}
\end{equation}
with certain dimensionless  constants $ f_i $  . With eq. (5.31)
we 
solve  eq. (5.22) and insert the result into the metric  (5.3). 
To 
simplify    the   expressions,    we   perform   the   coordinate
transformation   $   t  = l \cdot \tau  $    .  Then  the  metric
describing the Starobinsky inflation reads
\begin{equation}
ds\sp 2  = l\sp 2 [ d\tau \sp 2
- \exp(- \tau \sp 2 /6 )  \vert  \tau  \vert  \sp{-1/3}
 \sum \sp{\infty } _{i=0}
q_i \tau \sp{-2i} (dx \sp 2 + dy \sp 2 + dz \sp 2 ) 
]
\end{equation}
where  $  q_0   = 1 $ and the other $ q_i  $   are  certain  real
constants.  Metric  (5.32) gives a good presentation in the
region  
$ - \infty < \tau  \ll - 1 $.  One can see that it would not have
been  found  by a simple guess,  e.g.,  by a Fourier  or  Laurent
sequence  in $ t$ or so.  How to come from (5.29) to  the 
analogous 
expressions  for the $ f_i $  and the $ q_i $ is  straightforward
analysis,  and  the convergence of the sequences can  be  proved;
from the line after eq. (5.29) it becomes at least quite
plausible. 
  In  the presentation (5.32) it is not immediately clear  that 
this is inflation; for $ \tau  \ll -1 $, the parabola $ - \tau
\sp 2 /6 
$ is  an  almost linearly increasing function,  so that the 
cosmic 
scale  factor  is almost exponentially  increasing,  because  the
other terms do not essentially change the picture. 
         Let us now come the main question here: what happens for
$  l  \longrightarrow  0  $   ?   In  metric  (5.32),  all 
metric 
coefficients  can  be  developed  into powers of $  l\sp  2  $  ,
moreover, they are quadratic polynomials in $ l $. However, for $
l   \longrightarrow  0  $,   the  metric  degenerates.   This  is
essentially the argument of Simon [2], that Starobinsky inflation
is not selfconsistent in semiclassical gravity.  One should  look
whether this effect depends on the special coordinates chosen. To
this  end we go back to synchronized coordinates $ t = \tau  l 
$.  Then  the  factor   $ \exp (-t\sp 2 /6l\sp 2 )$   brings  the
problem  (besides  the  third root of $l$ in  the  next  factor),
whereas the further sequence is a sequence in $ l\sp 2 $  .  This
is  in  agreement  with  the  fact,  that  for  $  l  =  0$,  the
corresponding  field equation has only the flat Minkowski
 space-time as solution. 
The  recent view of the Starobinsky model can be  found  e.g.  in
refs. [41], [42] and a more geometrically oriented review in 
[43].
\bigskip
\bigskip
\section{Sixth and higher order equations}
\setcounter{equation}{0}
In  this  section,  we consider gravitational field equations  of
order higher that fourth;  this is mainly done to 
show,  how the fourth order Starobinsky model is situated between
the Einstein theory and the sixth and higher order ones.  
\subsection{ The Newtonian limit}
\setcounter{equation}{0}
The  Newtonian  limit  is the slow-motion  approximation  of  the
linearized field equation.  In this limit, the fourth order field
equation following from (5.19) becomes tractable.  For a $ 
\delta 
$-source of mass $  m $ one gets
\begin{equation}
ds \sp 2 = (1 - 2 \Phi ) dt \sp 2 - 
(1 + 2 \theta )( dr \sp 2 + r \sp 2 d \Omega \sp 2 )
\end{equation}
where $  d \Omega \sp 2  $ denotes the metric of the unit $ S \sp
2 $,
\begin{equation}
\Phi     =    \frac{m}{r}[1    - \frac{4}{3}    \exp(-r/k)    + 
\frac{1}{3}\exp(-r/l) ]        
\end{equation}
 see [32], and
\begin{equation}
\theta  = \frac{m}{r}[1 - \frac{2}{3} \exp(-r/k) -  
\frac{1}{3}\exp(-r/l) ]        
\end{equation}
see [15]. It is essential to observe that the solutions (6.2,
6.3) 
are unique. One should notice: Inspite of the higher order of the
differential equation one needs the same restriction (namely, the
vanishing of $ \Phi $  and $ \theta $ as $r$ tends to infinity) 
to 
get  a  unique  Newtonian  limit.  We have  considered  the  same
question  for  a  class  of  gravitational  field  equations   of
arbitrary  high  order  and  got the same  result [21]  for  the 
tachyonic-free case. We used
\begin{equation}
L=\frac{R}{2} - \frac{R}{12} \quad  \sum\sp{p}_{i=0} \quad 
\sum_{0 \le j_0 < j_1 < \dots < j_i \le p} \quad 
\prod \sp i _{m=0} l \sp 2 _{j_m} \quad 
\Box \sp i R
\end{equation}
where  $  p \ge 0 $ and $ 0 < l_0  < l_1  < \dots < l_p  $    are
characteristic lengthes and $ \Box $     denotes the
D`Alembertian. 
(For comparison:  eq. (5.19) with $ k = 0 $ and eq. (6.4) with $
p = 
0 $ coincide.) For eq. (6.4) the Newtonian limit gives (6.1) with
\begin{equation}
\Phi = \frac{m}{r} [1 + \frac{1}{3}
\sum \sp p _{i=0} (-1) \sp{i+p} 
\prod  _{j \ne i}  \vert  \frac{l \sp 2 _ j}{l \sp 2 _ i } - 1  
\vert  
\sp{-1} \exp(-r/l_i)]
\end{equation}
and  $ \Phi   +  \theta  = \frac{2m}{r} $.  For an extended  mass
distribution the result is the same because of the linearity, and
for the full nonlinear equations one can conjecture that at least
in the vicinity of flat space the result remains the same. 
     There  is  an essential point of departure from  the 
 Pauli-approach mentioned before and the calculations here: 
For $ L = R \sp 2 $  (ref.  [33]) and also for the other purely
quadratic 
Lagrangians (i.e.  linear combinations of the terms in  eq.
(5.17),  
see refs. [34, 35]) one does not get the correct Newtonian limit 
unless  one  adds the Einstein-Hilbert Lagrangian to the  action.
(Remark:  In [16] and [36] the same problem is considered with
the 
same Lagrangian but another variation (Palatini's one which gives
the same theory for the Einstein-Hilbert Lagrangian only),  i.e.,
independent  variation with respect to metric and  affinity;  the
result  agrees  not  only  with  respect to  the  fact  that  the
Einstein-Hilbert Lagrangian must be added,  but also with respect
to the general Newtonian plus Yukawa-type of the potential.) 
     Let  us end this section with a further point  of  departure
from Pauli's approach [26]:  He (and the authors of [37] and [38]
too)  required the outer solution to be developable in powers  of
$ m/r $.  But  then { \it only } the Schwarzschild solution
appears which  is 
definitely  { \it not } the outer solution for a point mass.  And
neither 
eq. (6.2)  nor (6.3) can be developed in powers of $1/r$.
\bigskip
\bigskip
\subsection{ Generalization of Simon's approach to higher order
gravity}
In the units chosen here ($ 8 \pi G = c = 1$) the Planck length 
$ l_{Pl} $ is related to Planck's constant via  $ \hbar  = 8  \pi
l_{Pl} \sp 2 $ .  So,  Simon's expansion [2, 11] into powers of $
\hbar $ 
is equivalent to an expansion into powers of $ l\sp 2 $ , where $
l  $  is a fixed length.  Let us take as example  the  Lagrangian
(6.4)  which was already considered in [22]  and [39] for $p = 1$
and in [23] for general $p$. Eq. (6.4) can be written as 
\begin{equation}
L=\frac{R}{2} - \frac{R}{12} \sum\sp{p}_{i=0}c_il\sp{2i+2}
\Box \sp i R
\end{equation}
with numerical constants $ c_i $ . We suppose $ c_p  \ne 0 $, and
(6.4)  leads to a field equation of order $ 2p + 4 $.  (We do not
need it in extent here;  one can find it in [23],  eq. (8).)  For
simplicity,  we consider the vacuum equations only.  We show that
by the method of Simon,  the order can be reduced to $2p + 2 $ as
follows:  We  suppose  (6.4) to be the truncation of an  infinite
sequence, and so it is valid only up to corrections of order $ O(
l \sp{2p+4}    )$.  Then we multiply the field equation following
from (6.4) by $ l \sp{2p+2} $ with the result that only the  term
from  the Einstein-Hilbert part of the Lagrangian  survives;  all
other  contributions can be subsumed to 
another $ O( l  \sp{2p+4})$. So we get
\begin{equation}
     0 =  R_{ij} l \sp{2p+2} +  O(l \sp{2p+4} ) 
\end{equation}
For $p = 0$,  this is just eq. (5.3) of ref. [11]. We can form
the 
covariant  derivatives  of  (6.7),  multiply it by $ R$  and 
form 
traces.  Then it is possible to add such a linear combination  of
these  equations to the field equation that,  up to terms of  the
order   $  O( l \sp{2p+4}    )$,  all terms stemming from $ c_p  
R      
\Box \sp p R $ in eq. (6.6) are compensated and the field
equation 
reduces  to  the order $2p+2$.  For $p = 0$ this  coincides  with
Simon's approach. 
     If  the higher order terms in the Lagrangian do not  contain
derivatives  of  the  curvature,  then the field equation  is  of
fourth order in each step;  this has been analysed in [12]; 
there 
it  is  also  mentioned  that  Starobinsky  inflation  remains  a
consistent  solution  if one interprets fourth order  gravity  as
classical theory and not as semiclassical one. This point of view
(see  also  [13])  is compatible with Stelle's  result [40]  that
fourth order gravity,  if taken as classical theory, becomes - in
contrast to Einstein's theory -  renormalizable. 
     But  here we have chosen an example where the order  of  the
differential  equation  is  increased  step  by  step.  The  next
question  which  is  interesting  for $ p >  0$  is  whether  the
procedure can be repeated such that the order can be reduced from
$ 2p+2 $ to  even  lower order,  one should expect that it  must 
be  
order  2 at the end.  The simplest non-trivial example is $p = 
1$, 
where,  after  the  first step  described  above,  the  following
fourth order equation appears 
\begin{equation}
  0  =  R_{ij} - \frac{R}{2}g_{ij}  - \frac{l \sp 2}{3} [ \Box  R
\,  g_{ij} 
- R_{;ij}  + R \, R_{ij}  - \frac{R \sp 2}{4} g_{ij}  ] +
    O(l \sp 6 ) 
\end{equation}
where  the  semicolon  denotes  the  covariant  derivative.   The
essential difference between eq. (6.8) and eq. (1.1) of [11] is
now
the power (here $ O(l \sp 6 )$, there $ O(l \sp 4 ))$ of the 
remainder. The trace of (6.8) reads
\begin{equation}
     0 =  R + l \sp 2  \Box R +  O(l \sp 6 ) 
\end{equation}
We apply $ l\sp 2 \Box $     to eq. (6.9) and get
\begin{equation}
     0 = l \sp 2 \Box R + l \sp 4 \Box \Box R +  O(l \sp 6 ) 
\end{equation}
The same done with (6.10) yields
\begin{equation}
     0 = l \sp 4 \Box \Box R +  O(l \sp 6 )                     
\end{equation}
The sum of eqs. (6.9) and (6.11) minus eq. (6.10) yields
\begin{equation}
     0 = R +  O(l \sp 6 )                     
\end{equation}
Similarly  one can handle the trace-free part of eq. (6.8).  So
we 
have  shown that (at least this type of) sixth order gravity  can
be brought to second order by Simon's approach. 
But  one should mention that we have,  as Simon  did,  made  such
assumptions  that the application of $l\sp 2\Box$ does not change
the  power  of  the  general  remainder.  This  is  a  consistent
assumption because  $l\sp 2\Box$ is a dimensionless operator.
 By inclusion of matter,  one gets then covariant derivatives  up
to  the  fourth  one of the energy-momentum  tensor  (instead  of
second  derivatives found in [11],  eq. (5.4)).  The
corresponding 
calculation is straightforwardly done, so we do not write out the
formulae. (Also,  they are not so  essential  here, 
because  in  regions where the higher order terms are dominant, 
one  usually  believes that matter is not yet essential  for  the
dynamics.)  Let us sketch them: The l.h.s. of eq. (6.7) becomes,
in 
analogy to eq. (5.3) of [11], 
$\kappa l\sp{2p+2}(T_{ij}-\frac{1}{2}Tg_{ij})$,  where $T_{ij}$
is 
the  energy-momentum tensor and $T$ its trace.  The l.h.s. of 
eq. 
(6.8)  gets the form $\kappa [T_{ij} + l\sp 2( \Box T  +$ 
similar 
terms)], and then two further covariant derivatives to the l.h.s.
appear similar as those one in eqs. (5.4)/(5.5) of ref. [11].
\bigskip
\bigskip
\section{Discussion}
The Starobinsky model goes back to early ideas of J. B. Zeldovich
and A.  D.  Sakharov,  see e.g.  ref. [44], where the addition of
higher  curvature  terms  to the Einstein  - Hilbert  action  was
intended  to  mimic quantum gravitational effects;  it was  hoped
that these terms can prevent the initial singularity.
Another approach can be found in  ref.  [45],  where  the 
stress  tensor  renormalization of quantized matter fields  in  a
classical  background  metric lead to curvature squared terms  in
the effective action with spin-dependent calculable  coefficients
in front of them. 
A  third approach was performed by Stelle [40],  who showed  that
the  Lagrangian  (5.19)  leads  to  a  renormalizable  theory  of
gravity; the coefficients in front of the curvature squared terms
are not calculable, but should be measured. 
We distinguished these three approaches explicitly,  because they
are often mixed.
\bigskip
The  Starobinsky  model follows from the  Lagrangian  eq.  (5.1),
which  coincides  with  eq.  (5.19) if $k = 0$.  This  is  not  a
renormalizable  theory  of gravity,  and it shares this  property
with Einstein's General Relativity Theory (GRT). 
One  instability of the theory following from  eq.  (5.19)  comes
from  the  fact,  that for $k\sp 2 <0$,  tachyons appear and  for
$k\sp 2>0$, ghosts appear (the latter are particles with negative
kinetic energy). The Starobinsky model, however, contains neither
tachyons nor ghosts. 
A  further instability can appear if there is no minimum  of  the
total  energy of a given local system.  In Einstein's GRT this is
prevented by the well-known positive energy theorem,  whereas eq.
(5.19)  with $k\ne 0$ allows an analogous theorem only in a  very
restricted sense,  cf.  ref.  [46]. For the theory following from
eq.  (5.1), however, a positive energy theorem, analogous to that
one  in GRT,  is valid,  see [47].  It needs as  only  additional
presumption  that  $R<3l\sp{-2}$.  This represents  no  practical
restriction   because  $l$  is  microscopically  small  and   the
inflationary   period  of  cosmic  evolution  is  connected  with
negative  values of $R$.  Connected with this fact is  the  point
discussed in sct.  3: requiring minimality of the action does not
rule out fourth order gravity. 
A  third  instability could occur if one looks at  eq.  (5.1)  as
perturbation  of  Einstein's  GRT,  $l$ playing the role  of  the
smallness parameter. This is, of course, a singular perturbation,
and  usually,  one would expect quickly increasing solutions  to 
appear.  In  general,  this takes place,  but under  the  special
circumstances met here, this does not happen. This has its origin
in  the special kind of nonlinearity of the singular differential
equation  (5.22):  it  has  the property that  for  each  initial
condition with $h>0$ (i.e.  initially, the universe expands), the
system can be integrated up to infinite time, and there it tends 
to  the corresponding solution of Einstein's GRT.  (We made  this
explicit  here  in sct.  5.3. because of  statements 
found in refs. [17, 18] which seem to contradict this, but 
in fact, only use another notion of instability.)
  This  regular 
behaviour  of  the  solutions can also be seen in  the  Newtonian

limit,  see  sct.  6.1.  If  one looks at  the  solutions  eqs. 
(6.2)/(6.3)  one can see that they converge to the  corresponding
Newtonian potential as $k,  l \longrightarrow 0$ but they  cannot
be  developed into powers of $k$ and $l$.  So the problem of  the
superfluous  degrees of freedom can be solved by stating that  in
the  weak-field  region,  the  coefficients of  these  terms  are
unobservably  small.  Another  way to deal with  the  superfluous
degrees of freedom is carried out by Simon in [2, 10]. 
 Also in [1], page 
408 there it is pointed out,  that the Starobinsky model is not
more 
unstable than Einstein's theory itself.
These  stability statements are all compatible.  To see this, one
has  to  remember that for initially  contracting  perturbations,
both Einstein's theory and fourth order gravity yield a big bang-
type instability after finite time. 

The  instability  appearing  from the fact that  a  fourth  order
equation  can be brought to a Hamiltonian with indefinite kinetic
energy,  see  refs.  [1,  4], was analyzed in detail in sct. 2.
We 
showed   at  some  typical  examples  that  this  can   lead   to
instabilities,  but  it  need not to do so.  We proposed  another
general   approach to bring a fourth order theory to  Hamiltonian
form in sct.  4.  The advantages of our approach are also  listed
there. The Hamiltonian form of the theory is needed to deduce the
Wheeler de Witt equation of the system.  In sct.  5.1. we made it
for  the  high-curvature  limit - and there the Wheeler  de  Witt
equation  could  be solved in closed  form,  eq.  (5.16).  It  is
planned  to  make the analogous calculations for  less  symmetric
space-times and the theory including the $R$-term,  i.e., for the
Lagrangian (5.1).  For the interpretation of them cf.  e.g.  ref.
[48].  But for the problem discussed here one only needs the sign
of  the kinetic energy in the Hamiltonian formulation;  it is the
same  as the signature of the superspace metric as is clear  from
eq.  (5.12). In ref. [49], the following was shown: The signature
$S$ (=number  of  negative  eigenvalues)  of  the  superspace 
metric 
leading  to the Wheeler de Witt equation following from
 Einstein's GRT equals $S=1 + s(n-s)$,  where $n$ is the 
dimension 
of  the spatial part (usually $=3$) of the space-time metric  and
$s$  its signature ($s=0$ both for the Lorentzian as well as  for
the  Euclidean  signature  of  the  underlying 
$(n+1)$-dimensional 
manifold).  So  $S=1$  for Einstein's GRT and  usual  signature, 
which  has the consequence that the Wheeler de Witt equation is a
normal hyperbolic wave equation. What it essential here: The fact
that  for fourth order gravity the Hamiltonian formulation  leads
to  an  indefinite  kinetic energy (superspace  metric  signature
equals 1) is a property which it has in common with GRT.
A  discussion  of the $R+R\sp 2$-theory  in  connection  with 
topological defects can be found in ref. [50].
Let  us finally make some remarks what happens if one  adds  some
higher  order  terms,   e.g.  those  one  discussed  in  sct.  6,
especially  the  Lagrangian (6.4) with $p \ge 1$ leading  to  the
order  of  the differential equation $\ge 6$.  Then the  problems
become  more serious.  The superspace metric gets signature  $\ge
2$,  so  the  Wheeler  de  Witt  equation  is  no  more  normally
hyperbolic. The  conformal transformation to  Einstein's  theory 
with  several  scalar  fields (see [22] for $p=1$  and  [23]  for
$p>1$), leads always to ghosts. The hope that sixth order gravity
naturally leads to models with double inflation is not fulfilled,
see  ref.  [39].  It is  unclear yet whether eighth order gravity
(partial  results can be found in ref.  [51] - further work is in
progress)  can  solve  these 
problems.  So  the  proposal by Simon [2] to reduce fourth  order
gravity seems practicable to be generalized (as we did in sct. 6)
to reduce sixth and higher order models to second order.
 
{ \it Acknowledgements.}  Discussions  with  W.-M.  Suen  and  A.

Berkin 
(during  the  MG 6 Conference,  Kyoto 1991),  with  U.  Kasper 
(after  his  lectures  at Potsdam university  in  1992/93),  and 
the 
advises of the referee helped  me  to 
understand  the  problem.
    Further,   I  thank  V.   M\"uller,  A. 
A. Starobinsky, S. Reuter, M. Rainer, S. Kluske, J. Audretsch, A.
Economou, J. Kurths and the members 
of the Rome-group leaded by F. Occhionero for valuable comments.
I  am  grateful  to  M.  Mohazzab  for  his  lecture  at  Potsdam
university,  especially for 
his  hint to ref.  [1].  Financial support from KAI e.V.  Berlin,

contract Nr. 015373/E and from DFG Bonn, contract Nr. Schm 
911/5-1,  are gratefully acknowledged.  
\bigskip
\bigskip
{ \it References}
[1] D. Eliezer, R. Woodard, Nucl. Phys. { \bf B 325}, 389 (1989).

[2] J. Simon, Phys. Rev. {\bf D 45}, 1953 (1992).
[3] M. Ostrogradski, Memoires Academie St. Petersbourg Series{\bf
 VI} vol.  {\bf 4},  385 (1850), especially pages 395 and 514.
 
[4] T.  Damour,  G.  Sch\"afer,  J.  Math.  Phys.  {\bf 32}, 127 
(1991); R. Hill,  J. Math. Phys. {\bf 8}, 1756 (1967); E. Kerner,
 J. Math. Phys. {\bf 6}, 1218 (1965).
[5]   A. A. Starobinsky, Phys. Lett.  { \bf B 91}, 99 (1980).
[6]  H.-J.  Schmidt,  Minimal,  not  only stationary  action  for
$R+R\sp 2$ theories of gravitation,  Preprint PRE-ZIAP  85-01, 
10 
pages, Potsdam 1985, unpublished.
[7] U. Kasper, Class. Quant. Grav. {\bf 10}, 869 (1993).
[8]  H.-J.  Schmidt,  Classical  mechanics with  lapse,  Potsdam
preprint Nr. 93/10, September 1993, submitted to J. Math. Phys.
[9] S. Reuter, H.-J. Schmidt, On the Wheeler de Witt equation for
homogeneous cosmological models,  Preprint Universit\"at  Potsdam
92/11,  1992, appeared in:  Diff. Geometry and Applic., Proc.
Conf. Opava, 
Ed.: O. Kowalski and  D. Krupka, 1993, p. 243. 
[10] J.  Simon,  Proc. Sixth M. Grossmann Meeting Kyoto 1991,
Ed.:      
H. Sato, WSPC Singapore 1992, p. 1246.
[11] L. Parker, J. Simon, Phys. Rev.  { \bf D 47}, 1339 (1993).
[12] A. Liddle, F. Mellor, Gen. Relat. Grav.  { \bf 24}, 897
(1992).
[13] P.  Steinhardt,  Proc. Sixth M. Grossmann Meeting Kyoto
1991,      
Ed.: H. Sato, WSPC Singapore 1992, p. 269.
[14] R.  Schimming, H.-J. Schmidt, NTM-Schriftenr. Gesch.
Naturw.,      
Techn., Med., Leipzig  { \bf 27}, 41 (1990).
[15] H.-J. Schmidt, Astron. Nachr.  { \bf 307}, 339 (1986).
[16] V. Hamity, D. Barraco, Gen. Relat. Grav.  { \bf 25}, 461
(1993).
[17] W.-M. Suen, Phys. Rev. Lett.  { \bf 62}, 2217 (1989).
[18] W.-M. Suen, Phys. Rev.  { \bf D 40}, 315 (1989).
[19] W.-M. Suen, P. Anderson, Phys. Rev.  { \bf D 35}, 2940
(1987).
[20] V. M\"uller, H.-J. Schmidt, Gen. Relat. Grav.  { \bf 17},
769 (1985).
[21] I. Quandt, H.-J. Schmidt, Astron. Nachr.  { \bf 312}, 97
(1991).
[22] S.  Gottl\"ober,  H.-J. Schmidt, A. Starobinsky, Class.
Quant. 
     Grav.  { \bf 7}, 893 (1990).
[23] H.-J. Schmidt, Class. Quant. Grav. { \bf  7}, 1023 (1990).
[24] A. Vilenkin, Phys. Rev.  { \bf D 32}, 2511 (1985).
[25] H.  Weyl, Gravitation und Elektrizit\"at, Sitzungsber.
Preuss.      
Akad. d. Wiss. Berlin  { \bf 1}, 465 - 480 (1918).
[26]  W.  Pauli,  Merkurperihelbewegung und Strahlenablenkung  in

Weyls  Gravitationstheorie,   Ber.  d.  Deutsch.  Physikal.  Ges.

 { \bf 21}, 742 - 750 (1919). 
[27]  C.  Lanczos,  Elektromagnetismus als nat\"urliche
Eigenschaft      
der Riemannschen Geometrie, Zeitschr. f. Physik  { \bf 73}, 147
(1932).
[28] G. t'Hooft, M. Veltman, Ann. Inst. H. Poincar\'e { \bf 20},
69 (1974).
[29] H.-J. Schmidt, Europhys. Lett.  { \bf 12}, 667 (1990).
[30] H.-J. Schmidt, Astron. Nachr.  { \bf 309}, 307 (1988).
[31] H.-J. Schmidt, Astron. Nachr.  { \bf 308}, 183 (1987).
[32] K. Stelle, Gen. Relat. Grav.  { \bf 9}, 353 (1978).
[33] E. Pechlaner, R. Sexl, Commun. Math. Phys.  { \bf 2}, 165
(1966).
[34] H.-J. Treder, Ann. Phys. (Leipzig)  { \bf 32}, 383 (1975).
[35]  P.  Havas,  Gen.  Relat.  Grav.   { \bf 8}, 631 (1977), 
see also H. 
     Goenner and P. Havas, J. Math. Phys.  { \bf 21}, 1159
(1980).
[36] G. Stephenson, J. Phys. A (Math. Gen.)  { \bf 10}, 181
(1977).
[37] B. Fiedler, M. G\"unther, Astron. Nachr.  { \bf 307}, 129
(1986).
[38] R. Pavelle, Phys. Rev. Lett.  { \bf 40}, 267 (1978).
[39]  L.  Amendola,  A.  Battaglia  Mayer,   S.  Capozziello,  S.
Gottl\"ober,  V.  M\"uller,  F. Occhionero, H.-J. Schmidt, Class.
Quant.      Grav. { \bf  10}, L43 - L47 (1993).
[40] K. Stelle, Phys. Rev.  { \bf D 16}, 953 (1977).
[41] S.  Gottl\"ober,  V. M\"uller, H.-J. Schmidt, A. A.
Starobinsky,      
Int. J. Mod. Phys.  { \bf D 1}, 257 - 279 (1992/93).
[42] S.  Capozziello,  F.  Occhionero,  L. Amendola, Int. J. Mod.
Phys.  { \bf D 1}, 615 - 639 (1992/93).
[43] H.-J. Schmidt, Fortschr. Phys. { \bf  41}, 179 - 199 (1993).
[44] A. D. Sakharov, Dokl. Akad. Nauk SSSR, {\bf 177}, 70 (1967).
[45]  N.  Birrell,  P.  Davies,  Quantum fields in curved  space,
Cambridge University Press 1982.
[46] H.-J. Schmidt, Ann. Phys. (Leipz.) {\bf 44}, 361 (1987).
[47] A. Strominger, Phys. Rev. {\bf D 30}, 2257 (1984).
[48] C. Kiefer, Phys. Rev. {\bf D 47}, 5414 (1993).
[49] H.-J. Schmidt, Diff. Geometry and Applic., Proc. Conf. Brno,
Ed.: D. Krupka, WSPC Singapore 1990, p. 405.
[50] J.  Audretsch,  A.  Economou, C. Lousto, Phys. Rev. {\bf D
47}, 
3303 (1993).
[51] A.  Battaglia Mayer,   H.-J.  Schmidt, Class. Quant. Grav. {
\bf  10}, 2441 (1993).
\bigskip
{\large Erratum}
\bigskip
In sct. V C of ref. [1], ''Starobinsky inflation
as a power series'', the calculations are correct,
but the interpretation of the solution as a 
non--singular one is wrong.
\bigskip
To elucidate the origin of that error we give two
lemmata. The coordinates $t, \ x, \ y, \ z$ shall cover all 
the reals, and $a(t)$ shall be an arbitrary strictly 
positive monotonously increasing smooth function defined 
for all
 real values $t$. (''smooth'' denotes 
$C^{\infty}$-differentiable.) Then it holds
\bigskip

{\bf Lemma 1}: The Riemannian space defined by 
$$ds^2= dt^2+a^2(t)(dx^2+dy^2+dz^2)$$
 is geodesically complete.
\bigskip
\noindent 
This is well-known and easy to prove; however, on the other hand
it holds
\bigskip
{\bf Lemma 2}: The Pseudoriemannian space defined by 

\begin{equation}
ds^2= dt^2-a^2(t)(dx^2+dy^2+dz^2)
\end{equation}
 is light-like geodesically complete iff 
\begin{equation}
\int_{- \infty}^0 \ a(t) \ dt \ \ = \ \ \infty
\end{equation}
\bigskip
\noindent
(''iff'' denotes ''if and only if''.)
The proof is straightforwardly done by considering light--like
 geodesics in the $x-t$-plane. Moreover,  
 lemma 2 remains valid if we replace 
''light--like geodesically complete'' by 
''light--like and time-like geodesically complete''.
(Lemma 2 seems to be unpublished up to now.)

\bigskip
\noindent
Let us now turn to the scope of this erratum: For the 
Lagrangian [1, eq.(5.1)]
$$
L = \left( \frac{R}{2} \ - \  \frac{l^2}{12} \, R^2 \right) 
\ \sqrt{-g}  \qquad {\rm where} \  l > 0
$$
\noindent 
one gets a fourth-order field equation; one of its solutions
is described in [1, eq.(5.32)]. In the region
$t \ll - l$, that solution can be approximated by
eq. (1) with 
\begin{equation}
a(t) \ = \ \exp( - \frac{t^2}{12 l^2} )
\end{equation}
However, this solution does not fulfil the
condition eq. (2). Therefore, by lemma 2, 
Starobinsky inflation does not 
represent a light--like geodesically complete cosmological model
as has been frequently stated in the literature, including
in ref. [1]. 
\bigskip
To prevent a further misinterpretation let me
reformulate as follows: Inspite of the fact that the Starobinsky 
 model is regular (in the sense that $a(t) > 0$ for arbitrary
values of synchronized time $t$), every past--directed 
light--like geodesic terminates in a curvature
singularity (i.e.,  $\vert R \vert \longrightarrow \infty$)
 at a finite value of its affine parameter.
Therefore, the model is not only geodesically
incomplete in the coordinates chosen, moreover, it also
fails to be a subspace of a complete one. 

\bigskip
\noindent
Let me add two remarks:  
\medskip
1.: Eq. (1) with $a(t) = \exp(H t)$, $H$ being a positive
constant, is the inflationary de Sitter space--time. 
According to lemma 2, it is also incomplete. 
However,  contrary to the Starobinsky model, 
it is a subspace of a complete space--time.
\medskip
2.: This erratum has no further consequences for the 
Starobinsky model. 
    
\bigskip
\noindent
[1] H.-J. Schmidt, Phys. Rev. D {\bf 49}, 6354 (1994).
\end{document}